\documentclass[prl,twocolumn,showpacs,floatfix]{revtex4}
\usepackage{graphicx}
\usepackage{bm}
\usepackage{amsmath}
\usepackage{amssymb}

\begin{document}

\title{Magnetic field induced quantum phase transitions in the two-impurity Anderson model}
\author{Lijun Zhu and Jian-Xin Zhu}
\affiliation{Theoretical Division and Center for Nonlinear Studies, Los Alamos National Laboratory,
Los Alamos, New Mexico 87545, USA}
\date{\today}

\begin{abstract}
In the two-impurity Anderson model, the inter-impurity spin exchange interaction favors a spin singlet state between two impurities leading to the breakdown of the Kondo effect. We show that a local uniform magnetic field can delocalize the quasiparticles to restore the Kondo resonance. This transition is found to be continuous, accompanied by not only  the divergence of the staggered (antiferromagnetic) susceptibility, but also the divergence of the uniform spin susceptibility. This may imply that the magnetic field induced quantum phase transitions in Kondo systems are in favor of the local critical type. 
\end{abstract}
\maketitle

The study on quantum phase transitions and critical phenomena has been an extraordinarily active area of research in condensed matter physics and quantum field theory. One example which has been extensively studied in experiments is the magnetic quantum phase transition (QPT) in heavy fermion metals~\cite{Si10}. Two theoretical scenarios are suggested for this QPT: the spin-density-wave picture based on itinerant quasiparticles,  and the local quantum criticality mandating the localization of quasiparticles.  While it is commonly acknowledged that the competition between the onsite Kondo coupling and the intersite spin exchange interaction, namely, the Ruderman-Kittel-Kasuya-Yosida (RKKY) interaction, plays the determinant role, there are few theoretical methods which can handle them on an equal footing. However, this problem is well-defined in the two-impurity Anderson (or Kondo) model,  which presents such a competition effect in an exactly solvable way~\cite{Jayaprakash81,Jones87,Sakai89,Jones89,Affleck92,Sire93,Fye94,Gan95,ZhuSq10}. With the Kondo coupling, the impurity spin forms a Kondo singlet state with the spins of the conduction electrons and a quasiparticle resonance peak develops at the Fermi energy, which is described by the Kondo effect. When the inter-impurity spin exchange interaction is antiferromagnetic and strong enough, the two impurity spins tend to form a singlet by themselves, against the formation of Kondo singlets, leading to the localization of quasiparticles. As a result, the quasiparticle spectra have a ``pseudogap'' at low energies~\cite{Sakai89,ZhuSq10}. It is found that, the phase transition between the Kondo resonance state and the inter-impurity spin singlet state is continuous~\cite{Jones87}, accompanied by the divergence of the staggered (antiferromagnetic) spin susceptibility together with a discontinuous change of the spectral weight at the Fermi energy~\cite{Sakai89,ZhuSq10}. 

A magnetic field has been an essential experimental tuning parameter to investigate the magnetic properties and spin correlations of condensed matter physics, which is also relevant to the two-impurity Anderson model.  However, a detailed analysis on the magnetic field effect for this model, especially close to the characteristic scales of the two-impurity quantum critical point (QCP), is still lacking, which is the purpose of this study. While there are some existing theoretical studies~\cite{Simon05,Chung07,EMinamitani10} targeting the double quantum dot, they are limited to the cases with large magnetic fields, in which the physical properties follow the Zeeman splitting effect~\cite{Simon05,Chung07}. While it is known that a local staggered magnetic field can induce a QPT as it directly couples to the critical staggered spin fluctuations, it is not known the role of a local uniform magnetic field, which is usually applied in experimental studies~\cite{JCusters03}. 

In this Letter, we report the first observation of a magnetic field induced quantum phase transition in the two-impurity Anderson model from a numerical study. We find that  a local uniform magnetic field on the two impurities can drive a transition from the inter-impurity spin singlet state to the Kondo resonance state, leading to the delocalization of quasiparticles. We further show that this transition is {\it continuous}, accompanied by the abrupt change of the quasi-particle spectral weight at the Fermi energy and the divergence in staggered spin susceptibility. In sharp contrast with the two-impurity QCP at zero field, the uniform spin susceptibility is found to be also divergent at this magnetic-field-induced QCP.  The new observation is suggestive that the field-induced QCP in heavy fermion systems~\cite{JCusters03} does have the local nature, as advocated in recent QCP 
theories~\cite{QSi01}.

The Hamiltonian for the two-impurity Anderson model can be written as 
\begin{eqnarray}
H &=& \sum_{{\bf k}\sigma} \epsilon_{\bf k} c^\dag_{{\bf k}\sigma} c_{{\bf k}\sigma}
+  \sum_{{\bf k}\sigma,(i=1,2)} \left({V_{\bf k} \over \sqrt{N_c}} e^{i {\bf k}\cdot {\bf r}_i} 
c^\dag_{{\bf k}\sigma}f_{i\sigma} + h.c.\right) \nonumber \\
 &&+  \sum_{(i=1,2),\sigma} \epsilon_f f^\dag_{i\sigma} f_{i\sigma} +\sum_{(i=1,2)} U n_{f i\uparrow}n_{f i\downarrow} \nonumber \\
&& + I {\bf S}_{f1} \cdot {\bf S}_{f2}+ h(S_{f1z}+S_{f2z} ) \;, 
\label{eq:hamiltonian}
\end{eqnarray}
which describes two interacting local orbitals  $f_{i\sigma}$ (Anderson impurities) in hybridization with a non-interacting conduction electron band $c_{{\bf k}\sigma}$ with the strength $V_{\bf k}$ at each impurity site ${\bf r}_i$. $\epsilon_f$ and $U$ are the energy level  and onsite Coulomb interaction for the local orbitals, respectively. $I$ is a direct spin exchange interaction between two impurities. We here consider a uniform magnetic field $B$ which acts on the impurity spins only, while $h\equiv g\mu_B B$ has the dimension of energy. $g$, $\mu_B$ are Land\'{e} factor and Bohr magneton, respectively. In reality, an applied magnetic field acts on the conduction electrons as well.  But as along as the Zeeman energy is much smaller than the bandwidth, we can safely neglect this effect and only consider its effect on the impurity spins. This model has been shown~\cite{Jones87,Sakai89,ZhuSq10} to be equivalent to a two-impurity two-channel model, with degrees of freedom casted into the even ($e$) and odd ($o$) parity channels.  The local orbitals become $f_{e,o} = (f_1 \pm f_2)/\sqrt{2}$ and the hybridization functions for these two channels are $\Gamma_{e,o}(\omega)  =   (1/ 2 N_c) \sum_{\bf k} V^2_{\bf k} |e^{i{\bf k}\cdot {\bf r}} \pm e^{-i{\bf k}\cdot {\bf r}}|^2 \delta( \omega -\epsilon_{\bf k} )$, where ${\bf r} = ({\bf r}_1-{\bf r}_2)/2$. The inter-site spin exchange interaction can be generated by considering specific forms of $V_{\bf k}$ and $\epsilon_{\bf k}$ (RKKY interaction), or provided by the direct spin exchange term $I$.  We adopt the numerical renormalization group (NRG)~\cite{Bulla08} method with the complete-Fock-space NRG (CFS-NRG) method~\cite{Anders06} for calculations of dynamical quantities at zero temperature, including the spectral function $A_{f,p\sigma}=-\text{Im}G_{f,p\sigma}(\omega)$, the uniform and staggered spin susceptibilities $\chi_{u,a} = \langle\langle S_{1z}\pm S_{2z}; S_{1z}\pm S_{2z} \rangle\rangle$.  We notice that the CFS-NRG method is particularly suitable for the problem with a finite magnetic field, as the characteristic energy scale is in the intermediate energy range (close to the Kondo temperature). This has also been evidenced from a similar study on the single impurity model~\cite{Zhang10}. 

In our previous study on this model~\cite{ZhuSq10}, we have explicitly calculated a system with a well-defined two-impurity QCP (we refer the two-impurity QCP to this zero-field QCP in the following). Our results can be summarized as follows. We choose $\Gamma_{e,o}(\omega) = \Gamma_0$, for which no RKKY interaction is generated and the single-impurity Kondo temperature $T_K$ can be determined. We then add the direct spin exchange interaction $I$ (to simulate RKKY interaction) to tune the competition between the Kondo effect and the inter-impurity spin exchange interaction. Below a critical value $I_c$, the low energy properties are still due to the Kondo effect, $A_f(0) \approx 1/(\pi \Gamma_0)$ and $\chi'_u(0)\sim 1/T_K$, but $\chi'_a(0) \sim 1/T_F^*$ with $T_F^*$ the reduced (local) Fermi liquid temperature.  Above $I_c$, it is the inter-impurity spin singlet state with vanishing $A_f(0$).  It differs from the Mott gap in the fact that there are still finite spectral weights at low energies: $A_f(\omega) \sim \omega^2$ for $\omega <T_F^*$ and a Non-Fermi liquid form for $T_F^* < \omega <T_{sf}$, where $T_F^*$ and $T_{sf}$ (spin fluctuation scale) correspond respectively to the two energy scales $T_L$ and $T_H$ identified in our previous work~\cite{ZhuSq10}. For $\Gamma_0 = 0.045\pi D$, and $\epsilon_f = -U/2 = -D$, it is found that the single-impurity Kondo temperature $T_K = 1.0\times 10^{-3}D$ and the critical value $I_c \approx 0.0023464D \approx 2.3 T_K$. At $I_c$, there is a sudden change of the spectral weight at the Fermi energy, but the transition is still {\it  continuous}. This is evidenced by the uniformly vanishing $T_F^* \sim (I-I_c)^2$ with the divergence of the staggered spin susceptibility $\chi'_a(0)$. However, the uniform spin susceptibility $\chi'_u(0)$ remains finite through the transition. 

\begin{figure}[tbh]
\includegraphics[width=\columnwidth]{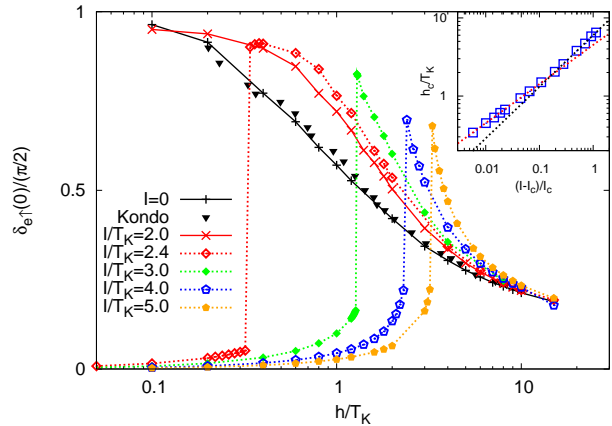}
\caption{(color online) The scattering phase shift $\delta_{e\uparrow}(0)$ as a function of $h$ for different values of $I$. The phase shifts at the Fermi energy for different channels and spins are the same~\cite{footnote1} due to the symmetries $A_{e\sigma}(\omega) = A_{o\sigma}(\omega)$ and $A_{e\uparrow}(\omega) = A_{e\downarrow}(-\omega)$. The data labelled with "Kondo" are subtracted from Costi's paper on the single-impurity Kondo model~\cite{Costi00}. The inset shows the critical value $h_c$ as a function of $I-I_c$ for $I>I_c$. Two dotted lines  are $h_c=4.5T_K (I/I_c-1)^{1/2}$ (red) and $h_c = 6.0 T_K (I/I_c-1)^{2/3}$ (black). }
\label{fig:phaseshift}
\end{figure}

We add a local uniform magnetic field $h$ to the above system to examine its effects. In Fig.~\ref{fig:phaseshift}, we show the results of the scattering phase shift $\delta_{p\sigma}(0)$ at the Fermi energy, determined from $A_f(0) = (1/\pi\Gamma_0)\sin^2\delta_{p\sigma}(0)$, as functions of $h$ for various values of $I$. For $I=0$, this is equivalent to a single-impurity Kondo problem and indeed our results are in agreements with those obtained for the single-impurity Kondo model~\cite{Costi00}. The scattering phase shift by the exact Bethe-Ansantz method is $\delta_h(0) \sim \pi/2 - h$ for $ h \ll T_K$ while $\delta_h(0)  \sim 1/\log(h/T_K)$ for $h \gg T_K$. The latter relation is in agreement for all finite $I$s but with $1/\log(h/T_{sf})$. For $I<I_c$, $\delta_h(0)$ is always finite and is enhanced from the single-impurity case for the same $h$. For $I>I_c$, $\delta_h(0)$ vanishes when $h=0$, characterizing the inter-impurity spin singlet state. A finite but small $h$ induces a small $\delta_h(0)$ or a small quasiparticle weight at the Fermi energy: it is found that $\delta_h(0) \sim h$ as verified by a log-log plot (not shown). When $h$ is increased to a critical value $h_c$, we observe a sudden jump of the phase shift from a tiny value to a large value of the order unity, which indicates a transition rather than a crossover between the inter-impurity spin singlet state and the Kondo resonance state. The relation between $h_c$ and $I-I_c$ is shown in the inset of Fig.~\ref{fig:phaseshift}. We find that $h_c \sim (I-I_c)^{1/2}$ for $(I-I_c)/I_c \ll 1$ and a noticeable deviation for $(I-I_c)/I_c > 0.1$. Such a deviation is also identified in $T_F^* \sim (I-I_c)^{\alpha}$ when $h=0$~\cite{ZhuSq10}.

\begin{figure}[tbh]
\includegraphics[width=0.9\columnwidth]{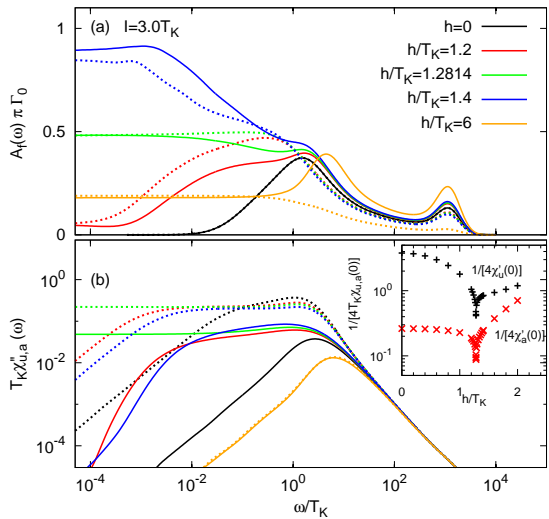}
\caption{(color online) Spectral functions $A_f(\omega)$ (a) and the imaginary parts of the uniform and staggered spin susceptibilities (b) as functions of energy for different values of $h$ in the inter-impurity singlet regime, $I= 3.0T_K$.  In (a), the solid and dotted lines represent the spin up and spin down degrees of freedom, respectively (the even and odd channels are the same). In (b), they respectively represent the uniform and staggered spin susceptibilities. The inset shows the real parts of the spin susceptibilities at zero energy (static) as functions of $h$. The inverse values are shown for convenience to be compared with the Kondo scale.}
\label{fig:varyhpg}
\end{figure}

\begin{figure}[tbh]
\includegraphics[width=0.9\columnwidth]{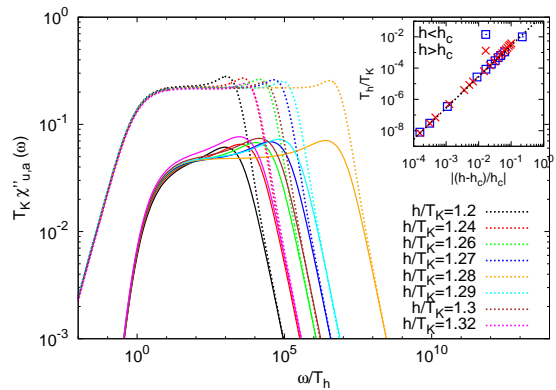}
\caption{(color online) Scaling behavior of the staggered spin susceptibility (dotted lines) for various values of $h$ near $h_c$. The rescaled uniform spin susceptibility (solid lines) is also shown. From the scaling, $T_h$ can be obtained and is plotted in the inset as a function of $|h-h_c|/h_c$. The line is a fitting $T_h/T_K = 0.28 | h/h_c-1|^2$. }
\label{fig:hcrscale}
\end{figure}

To show that it is a continuous phase transition, we present a detailed analysis for the $I=3.0T_K$ case. In Fig.~\ref{fig:varyhpg}, we show the results of the spectral functions $A_{f,p\sigma}(\omega)$, and the uniform and staggered spin susceptibilities $\chi_{u,a}(\omega)$. As the parity symmetry is not broken with a uniform magnetic field $h$, $A_{f,e\sigma}(\omega) = A_{f,o\sigma}(\omega)$. For a small magnetic field, for instance $h=1.2T_K$, the spectral weight at the Fermi energy is only slightly enhanced. Around $T_{sf}$, where spin fluctuations reach maximum,  the peak positions (on the positive energy range) of the spectral functions for the spin-up and spin-down degrees of freedom are different, with the energy difference roughly given by $h$. For a large magnetic field, for instance $h=6T_K$, which is bigger than either the Kondo scale or the RKKY scale, this is similar to the single-impurity Kondo model in the presence of a magnetic field.  Spin-up and spin-down resonance peaks are located respectively at positive and negative energies, and the energy difference (or the gap) is $2h$, which is the hallmark of Kondo resonance (for a non-interacting orbital, the gap from Zeeman splitting is $h$).  Around a critical value $h_c \approx 1.2814 T_K$, the spectral weight at the Fermi energy jumps from a tiny value to a value comparable with $1/(\pi\Gamma_0)$ but smaller. This is similar to the two-impurity QCP, except here the jump amplitude is smaller. The staggered spin susceptibility has the same behavior as well: it becomes divergent when $h\to h_c$. The significant difference lies in the uniform spin susceptibility: it is also divergent at $h_c$ in this field-induced QCP compared with finite $\chi_u(0) \sim 1/T_{sf}$ in the two-impurity QCP.  While we can subtract the energy scales from $1/[4\chi_{u,a}(0)]$ as determining $T_K$, a more reliable method to determine the low energy scale is from a scaling analysis, which is shown in Fig.~\ref{fig:hcrscale}, for $\chi''_a(\omega)$. Once the energy is scaled with a certain scale $T_h$ for different $h$, the low energy part of $\chi''_a(\omega)$ falls into a universal curve. Similar behavior can be observed in $\chi''_u(\omega)$, but its high energy part does not appear to scale (or not universal). We further plot the obtained $T_h$ as a function of $|h-h_c|$:  $T_h$ can be fitted as $T_h\sim |h-h_c|^2$ for both $h<h_c$ and $h>h_c$.  $T_h\to0$ as well as the divergence in both the staggered and uniform spin susceptibilities are clear evidences that this magnetic field induced QPT is continuous. 

\begin{figure}[tbh]
\includegraphics[width=0.9\columnwidth]{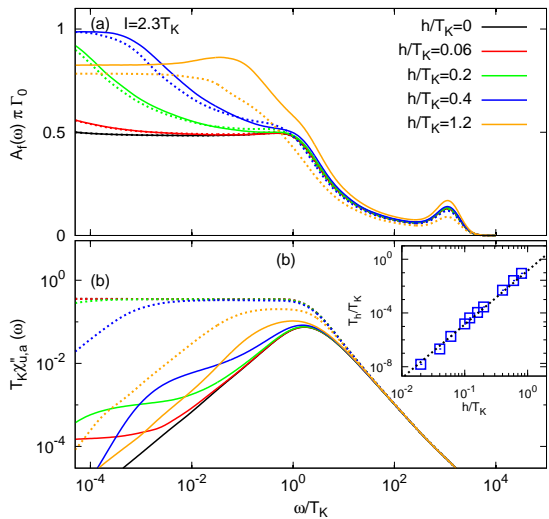}
\caption{(color online) Spectral functions $A_f(\omega)$ (a) and the imaginary parts of the uniform and staggered spin susceptibilities (b) as functions of energy for different values of $h$ in the two-impurity quantum critical regime, $I= 2.3T_K$. The line representations in (a) and (b) are the same as in Fig. 2. The inset shows $T_h$ as a function of $h$, which is obtained the same way as in Fig. 3. The line is a fitting $T_h/T_K = 0.18 (h/T_K)^4$.}
\label{fig:hqc}
\end{figure}

When $I\to I_c$, $h_c$ vanishes and the field-induced QCP merges with the two-impurity QCP at zero-field. We then need to understand why the divergence in $\chi_u$ vanishes. The results for the spectral functions and the spin susceptibilities for $I=2.3T_K$ are shown in Fig.~\ref{fig:hqc}. Indeed, any small $h$ induces the full Kondo resonance at the Fermi energy with the finite Fermi temperature $T_h$, which is fitted as $T_h\sim h^4$. We also observe enhancement of uniform spin fluctuations, which is manifested as a flat part above $T_h$, $\chi''_u(\omega) \sim C_h$. $C_h$ increases as $h$ increases. When $T_h$ vanishes as $h\to 0$, $C_h$ also vanishes. As a result, $\chi_u$ is not divergent. However, for any $I>I_c$, as $C_h$ remains finite, $\chi_u$ indeed diverges as $T_h$ vanishes when $h\to h_c$. The variation of $C_h$ also explains that $\chi''_u(\omega)$ does not scale for $\omega>T_h$ [cf. Fig.3]. 

The continuous QPT induced by a local uniform magnetic field in the two-impurity Anderson model was not predicted by either the conformal field theory~\cite{Affleck92} or the bosonization construction~\cite{Sire93,Gan95}. Compared with the two-channel QCP where $T_h \sim h^2$~\cite{Zarand08}, the analogy of the magnetic field in the two-channel Kondo impurity model is the staggered magnetic field in the two-impurity model, i.e., $h_s(S_{1z}-S_{2z})$, which directly couples to the critical staggered spin fluctuations. Our results suggest that the uniform magnetic field $h$ also couples effectively to the staggered spin fluctuations, as evidenced from the divergence of $\chi_a$. We can make the following statements on this coupling term based on our results. 1) The control parameter is modified as $I-I_c -a h^2$,  to be consistent with the exponents deduced from our numerical data in different regimes. In other words, the field-induced QCP is the same in nature as the two-impurity QCP at zero field. From another perspective,  for a given finite $h$, tuning $I$ can also lead to a QPT but the critical value of $I_c$ is shifted up. 2) It involves the uniform spin fluctuations as $h$ is directly coupled to. This also accounts for the divergence of the uniform spin susceptibility. In other words, this divergence is induced rather than the driving mechanism. Similar divergences are also observed in the non-local superconducting fluctuations and the current fluctuations between two impurities. 3) It vanishes or becomes irrelevant as $h$ vanishes, as necessary to explain the loss of divergence in $\chi_u$ at the two-impurity QCP. 
 
In a lattice generalization from the self-consistently solved two-site cluster, the uniform and staggered spin susceptibilities in this cluster correspond to the lattice spin susceptibility at momentum points $Q_0=(0,0,0)$ and $Q_{\pi}=(\pi,\pi,\pi)$ for a three-dimensional (3D) lattice. The divergence in $\chi_a$ relates to the antiferromagnetic instability near $Q_{\pi}$. However, in 3D, the spin density of states near $Q_\pi$ vanishes, unlike in 2D. The local spin susceptibility, which is a sum of contributions from all momentum points, is not divergent. This may relate to the spin-density-wave type of transitions.  If the transition is driven by the magnetic field, we learn from this study that the divergence at $Q_{\pi}$ can induce the divergence in $Q_{0}$ as well. The local spin susceptibility in this case diverges even for 3D. This corresponds to the local critical type of transitions. This might be related to the local quantum behaviors observed in YbRh$_2$Si$_2$ tuned by the magnetic field~\cite{Gegenwart02}.

\begin{acknowledgements}
We thank I. Affleck and C. M. Varma for helpful discussions. This work was supported by the National Nuclear Security Administration
of the U.S. DOE  at  LANL  under Contract No. DE-AC52-06NA25396, the U.S. DOE Office of Science, and the LDRD Program at LANL.
\end{acknowledgements}


\begin{thebibliography}{99}
\bibitem{footnote1} In practice, there is a small difference between $A_{e\uparrow}(\omega)$ and $A_{e\downarrow}(\omega)$ at the lowest energy point, which is due to the NRG method~\cite{Zarand08}. In our calculations, we find that the difference is in general a few percent and does not affect our conclusion on the phase transitions.
\end{thebibliography}
\end{document}